\documentclass{article}
\usepackage[preprint]{spconf} 
\usepackage{amsmath,amssymb,amsfonts}
\usepackage{graphicx,color}
\usepackage{textcomp}
\usepackage[dvipsnames]{xcolor}
\usepackage{hyperref}
\hypersetup{hidelinks=true}
\def\BibTeX{{\rm B\kern-.05em{\sc i\kern-.025em b}\kern-.08em
    T\kern-.1667em\lower.7ex\hbox{E}\kern-.125emX}}
\AtBeginDocument{\definecolor{tmlcncolor}{cmyk}{0.93,0.59,0.15,0.02}\definecolor{NavyBlue}{RGB}{0,86,125}}

\usepackage{etoolbox}

\makeatletter
\renewcommand\section{\@startsection{section}{1}{0pt}{1.5ex}{1ex}{\normalfont\Large\bfseries}}
\makeatother

\usepackage{balance}        
\usepackage{bm}			
\usepackage[per-mode=symbol]{siunitx}    		
\DeclareSIUnit{\dBm}{dBm}	
\usepackage[
    indexonlyfirst, 
]{glossaries}
\makenoidxglossaries%
\newacronym{2d}{2D}{two-dimensional}
\newacronym{3d}{3D}{three-dimensional}
\newacronym{ac}{AC}{alternating current}
\newacronym{am}{AM}{amplitude modulation}
\newacronym{aoa}{AoA}{angle of arrival}
\newacronym{afe}{AFE}{analog front end}
\newacronym{aod}{AoD}{angle of departure}
\newacronym{asic}{ASIC}{application specific integrated circuit}
\newacronym{ambc}{AmBC}{ambient BC}
\newacronym{awgn}{AWGN}{additive white Gaussian noise}
\newacronym{ar}{AR}{augmented reality}
\newacronym{bibc}{BiBC}{bistatic BC}
\newacronym{blue}{BLUE}{best linear unbiased estimator}
\newacronym{ce}{CE}{channel estimation} 
\newacronym{cir}{CIR}{channel impulse response}
\newacronym{cse}{CSE}{channel state estimation}
\newacronym{csi}{CSI}{channel state information}
\newacronym{crlb}{CRLB}{Cram\'er-Rao lower bound}
\newacronym{ct}{CT}{cosine transform} 
\newacronym{cjt}{CJT}{coherent joint transmission}

\newacronym{da}{DA}{data association}
\newacronym{dc}{DC}{direct current}
\newacronym{das}{DAS}{distributed antenna systems}
\newacronym{dli}{DLI}{direct link interference}
\newacronym{dl}{DL}{downlink}
\newacronym{ul}{UL}{uplink}
\newacronym{dm}{DM}{diffuse multipath}
\newacronym{dmc}{DMC}{diffuse multipath component}
\newacronym{dmimo}{D-MIMO}{distributed MIMO}
\newacronym{dtft}{DTFT}{discrete-time Fourier transform}
\newacronym{doa}{DOA}{direction-of-arrival}
\newacronym{dut}{DUT}{device under test}
\newacronym{ec}{EC}{European Commission}
\newacronym{eirp}{EIRP}{equivalent isotropically radiated power}
\newacronym{en}{EN}{energy neutral}
\newacronym{end}{END}{energy neutral device}
\newacronym{epu}{EPU}{edge processing unit}
\newacronym{epc}{EPC}{electronic product code}
\newacronym{esd}{ESD}{electrostatic discharge}
\newacronym{erp}{ERP}{equivalent radiated power}
\newacronym{etsi}{ETSI}{European Telecommunications Standards Institute}
\newacronym{evd}{EVD}{eigenvalue decomposition}
\newacronym{fa}{FA}{federation anchor}
\newacronym{fem}{FEM}{finite element analysis}
\newacronym{fdd}{FDD}{frequency-division duplexing}
\newacronym{fir}{FIR}{finite impulse response}
\newacronym{fss}{FSS}{frequency selective surface}
\newacronym{glrt}{GLRT}{generalized log-likelihood ratio test}
\newacronym{gsm}{GSM}{Global System for Mobile Communications}  %
\newacronym{eh}{EH}{energy harvesting}
\newacronym{ic}{IC}{integrated circuit}
\newacronym{id}{ID}{information decoding}
\newacronym{iid}{i.i.d.}{independent and identically distributed}
\newacronym{iot}{IoT}{Internet of Things}
\newacronym{ism}{ISM}{industrial, scientific and medical}
\newacronym{jfet}{JFET}{junction field effect transistor}
\newacronym{lis}{LIS}{large intelligent surface}
\newacronym{lmmse}{LMMSE}{linear MMSE}
\newacronym{los}{LoS}{line-of-sight}
\newacronym{ls}{LS}{least-squares}
\newacronym{lti}{LTI}{linear time-invariant}
\newacronym{mc}{MC}{Monte Carlo}
\newacronym{mimo}{MIMO}{multiple-input multiple-output}
\newacronym{miso}{MISO}{multiple-input single-output}
\newacronym{ml}{ML}{maximum likelihood}
\newacronym{mmwave}{mmWave}{millimeter wave}
\newacronym{mmse}{MMSE}{minimum mean square error}
\newacronym{mobc}{MoBC}{monostatic BC}
\newacronym{mosfet}{MOSFET}{metal-oxide semiconductor field effect transistor}
\newacronym{mn}{MN}{matching network}
\newacronym{mpc}{MPC}{multipath component}
\newacronym{music}{MUSIC}{MUltiple SIgnal Classification}
\newacronym{mrt}{MRT}{maximum ratio transmission}
\newacronym{mppt}{MPPT}{maximum power point tracker}
\newacronym{nfc}{NFC}{near-field communication}
\newacronym{nlos}{NLoS}{non-LoS}
\newacronym{olos}{OLoS}{obstructed LoS}
\newacronym{nmos}{nMOS}{n-channel metal-oxide semiconductor}
\newacronym{ofdm}{OFDM}{orthogonal frequency-division multiplexing}
\newacronym{ook}{OOK}{on–off keying}
\newacronym{ota}{OTA}{over-the-air}
\newacronym{nb}{NB}{narrowband}
\newacronym{nr}{NR}{new radio}
\newacronym{pc}{PC}{pilot count}
\newacronym{pce}{PCE}{power conversion efficiency}
\newacronym{pf}{PF}{particle filter}
\newacronym{pg}{PG}{path gain}
\newacronym{pl}{PL}{path loss}
\newacronym{pwm}{PWM}{pulse-width modulation}
\newacronym{rcs}{RCS}{radar cross section}
\newacronym{rf}{RF}{radio frequency}
\newacronym{rfid}{RFID}{radio frequency identification}
\newacronym{ris}{RIS}{reflective intelligent surface}
\newacronym{rms}{RMS}{root mean square}
\newacronym{rmse}{RMSE}{root mean square error}
\newacronym{rrmse}{RRMSE}{relative root mean square error}
\newacronym{rss}{RSS}{received signal strength}
\newacronym{rssi}{RSSI}{received signal strength indicator}
\newacronym{rw}{RW}{RadioWeave}
\newacronym{rv}{RV}{random variable}
\newacronym{s-parameter}{S-parameter}{scattering parameter}
\newacronym{sa}{SA}{synchronization anchor}
\newacronym{sar}{SAR}{specific absorption rate}
\newacronym{sbl}{SBL}{sparse Bayesian learning}
\newacronym{sdn}{SDN}{software-defined network}
\newacronym{srd}{SRD}{short-range device}
\newacronym{sinr}{SINR}{signal-to-interference-plus-noise ratio}
\newacronym{sir}{SIR}{signal-to-interference ratio}
\newacronym{siso}{SISO}{single-input single-output}
\newacronym{simo}{SIMO}{single-input multiple-output}
\newacronym{slam}{SLAM}{simultaneous localization and mapping}
\newacronym{snr}{SNR}{signal-to-noise ratio}
\newacronym{smc}{SMC}{specular multipath component}
\newacronym{svd}{SVD}{singular value decomposition}
\newacronym{si}{SI}{self interference}
\newacronym{sw}{SW}{spherical wavefront}
\newacronym{spa}{SPA}{sum-product algorithm}
\newacronym{tdd}{TDD}{time division multiplexing}
\newacronym{tdoa}{TDOA}{time-difference-of-arrival}
\newacronym{toa}{TOA}{time-of-arrival}
\newacronym{tosm}{TOSM}{through–open–short–match}
\newacronym{ue}{UE}{user equipment}
\newacronym{uhf}{UHF}{ultra high frequency}
\newacronym{ula}{ULA}{uniform linear array}
\newacronym{upa}{UPA}{uniform planar array}
\newacronym{ura}{URA}{uniform rectangular array}
\newacronym{uv}{UV}{unmanned vehicle}
\newacronym{vna}{VNA}{vector network analyzer}
\newacronym{wb}{WB}{wideband}
\newacronym{wpt}{WPT}{wireless power transfer}
\newacronym{wrt}{w.r.t.}{with respect to}
\newacronym{xets}{XETS}{cross exponentially tapered slot}
\newacronym{zf}{ZF}{zero-forcing}
\newacronym{csp}{CSP}{contact service point}
\newacronym{ecsp}{ECSP}{edge computing service point}
\newacronym{cmos}{CMOS}{complementary metal oxide semiconductor}
\newacronym{ask}{ASK}{amplitude shift keying}
\newacronym{fdma}{FDMA}{frequency division multiple access}
\newacronym{fsk}{FSK}{frequency shift keying}
\newacronym{ber}{BER}{bit error rate}
\newacronym{bp}{BP}{belief propagation}
\newacronym{pdf}{PDF}{probability density function}
\newacronym{psk}{PSK}{phase shift keying}
\newacronym{qam}{QAM}{quadrature amplitude modulation}
\newacronym{cw}{CW}{continuous wave}
\newacronym{papr}{PAPR}{peak-to-average power ratio}
\newacronym{bc}{BC}{backscatter communication}
\newacronym{ca}{CE}{carrier emitter}
\newacronym{bd}{BD}{backscatter device}

\newacronym{uwb}{UWB}{ultra-wideband}
\newacronym{bgmm}{BGMM}{Bayesian Gaussian mixture model}
\newacronym{gmm}{GMM}{Gaussian mixture model}
\newacronym{knn}{kNN}{k-Nearest-Neighbors}
\newacronym{ft}{FT}{Fourier transform}
\newacronym{dft}{DFT}{discrete Fourier transform}

\newacronym{adc}{ADC}{analog-to-digital converter}
\newacronym{np}{NP}{Neyman-Pearson}
\newacronym{pana}{PanA}{Panel A}
\newacronym{panb}{PanB}{Panel B}
\newacronym{p1}{P1}{Phase 1}
\newacronym{p2}{P2}{Phase 2}
\newacronym{pw}{PW}{planar wavefront}
\newacronym{i.i.d.}{i.i.d.}{independent and identically distributed}
\newacronym{pcsi}{PCSI}{perfect channel state information}
\newacronym{mrfh}{MRFH}{main RF harvester}
\newacronym{arfh}{ARFH}{auxiliary RF harvester}
\newacronym{mcu}{MCU}{microcontroller unit}

\newacronym{fim}{FIM}{Fisher information matrix}
\newacronym{efim}{EFIM}{equivalent FIM}
\newacronym{mva}{MVA}{master virtual anchor}
\newacronym{pa}{PA}{physical anchor}
\newacronym{pcrlb}{PCRLB}{posterior CRLB}
\newacronym{peb}{PEB}{positioning error bound}
\newacronym{ceb}{CEB}{clock error bound}
\newacronym{meb}{MEB}{mapping error bound}
\newacronym{pmva}{PMVA}{potential MVA}
\newacronym{va}{VA}{virtual anchor}
\newacronym{vm}{VM}{virtual mobile}

\usepackage{fontawesome}		
\usepackage{array, booktabs, xltabular}
\usepackage{colortbl}       
\usepackage{multirow}
\usepackage{enumitem}   
\usepackage[nice]{nicefrac}     	
\usepackage{textcomp}       

\usepackage{url}
\newcommand{\icon}[3]{\makebox(#2, #2){\textcolor{#3}{\csname fa#1\endcsname}}}	

\usepackage{caption}
\usepackage[labelformat=simple]{subcaption}

\usepackage{placeins}
\usepackage{tikzscale}		
\usepackage{tcolorbox}

\usepackage{tikz}
\usepackage{ifthen}

\usepackage[framemethod=TikZ]{mdframed}	

\usepackage{pgfplots}
  \tcbset{shield externalize} 
  \pgfplotsset{compat=newest}
  \usetikzlibrary{plotmarks}
  \usetikzlibrary{arrows.meta}
  
  \usetikzlibrary{arrows}
  \usetikzlibrary{mindmap,trees}
  \usetikzlibrary{positioning,spy}
  \usetikzlibrary{calc,fadings,decorations.pathreplacing}
  \usetikzlibrary{decorations.pathmorphing,decorations.text}
  \usetikzlibrary{datavisualization.polar}      
  \usepackage{pgfplotstable}
  \usepgfplotslibrary{polar}
  \usetikzlibrary{patterns}

  \usetikzlibrary{plotmarks}
  \usepgfplotslibrary{patchplots}
  \usepackage{grffile}
  \usepackage{amsmath}
  \usepackage{tikzscale}		
  \usepackage{tikz-3dplot} 
  \usetikzlibrary{positioning}
  \usetikzlibrary{arrows}
  \usetikzlibrary{fit,backgrounds}  

\pgfdeclarelayer{behindBackground}
\pgfdeclarelayer{background}
\pgfdeclarelayer{foreground}
\pgfsetlayers{behindBackground,background,main,foreground}   

\tikzset{%
  >=latex,
  inner sep=0pt,%
  outer sep=2pt,%
  mark coordinate/.style={inner sep=0pt,outer sep=0pt,minimum size=3pt,
  fill=black,circle}%
}
\newcommand{\externalizeFigures}{false}
\ifthenelse{\equal{\externalizeFigures}{true}}
{
  \usepackage{pgfplots}
  \pgfplotsset{compat=newest}
  \usepgfplotslibrary{external}
  \tikzexternalize[prefix=tikz/]
  \usepackage{shellesc}
}
{
\newcommand{\tikzexternaldisable}{}
\newcommand{\tikzexternalenable}{}
}

\newlength\figureheight
\newlength\figurewidth

\newcommand{\trajectoryLW}{1.0pt}     

\newcommand{\herm}{\mathsf{H}}

\newcommand{\trp}{\mathsf{T}}

\newcommand{\realset}[2]{ \mathbb{R}^{#1 \times #2}  }
\newcommand{\realsetone}[1]{\mathbb{R}^{#1}}

\newcommand{\complexsetone}[1]{ \mathbb{C}^{#1}  }

\DeclareMathOperator{\tr}{tr}

\DeclareMathOperator{\arctantwo}{arctan2}

\usepackage{amsthm}

\usepackage{algorithm2e}		
\RestyleAlgo{ruled}			
\SetKw{KwBy}{by}			%
\SetKwComment{Comment}{\% }{}	

\usepackage[hang,flushmargin]{footmisc}
\makeatletter
\newcommand{\algorithmfootnote}[2][\footnotesize]{%
  \let\old@algocf@finish\@algocf@finish
  \def\@algocf@finish{\old@algocf@finish
    \leavevmode\rlap{\begin{minipage}{\linewidth}
    #1#2
    \end{minipage}}%
    \vspace{-0.3cm} 
  }%
}

\definecolor{RDlightgreen}{RGB}{141 192 69}
\definecolor{RDgreen}{rgb}{0.3647, 0.4275, 0.2667}
\definecolor{RDdarkgreen}{rgb}{0.2196, 0.2196, 0.2196}
\definecolor{RDmaroon}{rgb}{.522,.22,.353} %

\definecolor{IEEEblue}{RGB}{0 98 155}
\definecolor{IEEElightblue}{RGB}{0 181 226}
\definecolor{IEEEturquoise}{RGB}{0 156 166}
\definecolor{IEEEred}{RGB}{186 12 47}
\definecolor{IEEEgreen}{RGB}{0 132 61}
\definecolor{IEEElightgreen}{RGB}{120 190 32}
\definecolor{IEEEorange}{RGB}{225 163 0}
\definecolor{IEEEyellow}{RGB}{255 209 0}
\definecolor{IEEEviolett}{RGB}{152 29 151}
\definecolor{IEEEdarkmaroon}{RGB}{134 31 65}

\definecolor{myblue}{rgb}{0.00000,0.60580,0.87650}  

\colorlet{AP1color}{IEEEblue}
\colorlet{AP2color}{IEEEred}
\colorlet{AP3color}{IEEEgreen}
\definecolor{AP4color}{rgb}{0.2,0.2,0.2}

\newcolumntype{C}{@{\hskip 0.075cm}c@{\hskip 0.075cm}}
\newenvironment{bmatrixs}
  {\left[\begin{array}{*{20}{C}}}
  {\end{array}\right]}

\newcommand{\elxAoD}[1]{\elevationAoD_{\scriptscriptstyle n,#1}^{\MVApair{s}{s'}}}             
\newcommand{\azxAoD}[1]{\azimuthAoD_{\scriptscriptstyle n,#1}^{\MVApair{s}{s'}}}               

\newcommand{\etarot}{\bm{\eta}^{\text{\tiny{TX,o}}}}

\newcommand{\eye}[1]{\mathbf{I}_{\scriptscriptstyle #1}}                
\newcommand{\diff}[0]{\mathrm{d}}       

\newcommand{\fc}{f_{\text{\tiny c}}}                                
\newcommand{\lightspeed}{\mathsf{c}}                                

\newcommand{\Nfrequency}[0]{ N_{\scriptscriptstyle\!f} }             
\newcommand{\Nantennas}[0]{ M }                                     
\newcommand{\Nantennasy}[0]{ M_y }                                  
\newcommand{\Nantennasz}[0]{ M_z }                                  
\newcommand{\stateSpace}[0]{\mathcal{S}_{\scriptscriptstyle\!\bm{\theta}}}                            
\newcommand{\Nparticles}[0]{ N_{\text{\tiny p}} }            

\newcommand{\MVApair}[2]{(\!#1,#2\!)}                               

\newcommand{\Pjss}[3]{\bm{P}_{\scriptscriptstyle #1}^{\MVApair{#2}{#3}}}                           

\newcommand{\origin}[0]{\bm{0}}                                     
\newcommand{\pitch}[0]{\theta^{\text{\tiny r}}_{\scriptscriptstyle j}}                     
\newcommand{\rotM}[0]{\bm{M}_{\scriptscriptstyle j}}                                       

\newcommand{\range}[4]{\bm{r}_{\scriptscriptstyle#1,#2}^{\scriptscriptstyle\MVApair{#3}{#4}}} 
\newcommand{\ranget}[4]{\widetilde{\bm{r}}_{\scriptscriptstyle#1,#2}^{\scriptscriptstyle\MVApair{#3}{#4}}} 
\newcommand{\rangepTX}[4]{\grave{\bm{r}}_{\scriptscriptstyle#1,#2}^{\scriptscriptstyle\MVApair{#3}{#4}}} 
\newcommand{\rangep}[4]{\acute{\bm{r}}_{\scriptscriptstyle#1,#2}^{\scriptscriptstyle\MVApair{#3}{#4}}} 
\newcommand{\rx}[0]{\acute{\bm{r}}_{\scriptscriptstyle x}}
\newcommand{\ry}[0]{\acute{\bm{r}}_{\scriptscriptstyle y}}
\newcommand{\rz}[0]{\acute{\bm{r}}_{\scriptscriptstyle z}}

\newcommand{\PiPerp}[2]{\bm{\Pi}_{\scriptscriptstyle \MVApair{#1}{#2}}^{\perp}}  

\newcommand{\observation}[2]{\bm{z}_{\scriptscriptstyle #1}^{\scriptscriptstyle(#2)}}   
\newcommand{\noise}[2]{\bm{w}_{\scriptscriptstyle #1}^{\scriptscriptstyle(#2)}}         
\newcommand{\dictEntry}[2]{\bm{\psi}_{\scriptscriptstyle j}^{\scriptscriptstyle \MVApair{#1}{#2}}}         

\newcommand{\PMVAposStacked}[1]{\overline{\bm{p}}^{\text{\tiny mva}}_{\scriptscriptstyle \,}}                   
\newcommand{\state}[1]{\bm{x}_{\scriptscriptstyle #1}}                                  
\newcommand{\covarianceEstimate}[1]{\widehat{\bm{P}}_{\scriptscriptstyle#1}}                
\newcommand{\pos}[1]{ \bm{p}_{\scriptscriptstyle #1} }                                  
\newcommand{\posEstimate}[1]{ \hat{\bm{p}}_{\scriptscriptstyle #1} }                    

\newcommand{\Rhate}[2]{\widehat{\bm{R}}_{\scriptscriptstyle #1,#2}^{\text{\scriptsize e}}} 

\newcommand{\particle}[2]{\bm{\theta}_{\scriptscriptstyle #1}^{\scriptscriptstyle(#2)}}     
\newcommand{\weight}[2]{w_{\scriptscriptstyle #1}^{{\scriptscriptstyle(#2)}}}          

\newcommand{\jacobian}{ \boldsymbol{J} }
\newcommand{\jacobMVAsb}[1]{ \jacobian_{\scriptscriptstyle n,j}^{\scriptscriptstyle \text{\tiny sb},s}}   

\newcommand{\elevation}{\theta}                                     
\newcommand{\azimuth}{\vartheta}                                    
\newcommand{\toffset}[1]{\epsilon_{#1}}                                                  

\newcommand{\elx}[1]{\elevation_{\scriptscriptstyle n,#1}^{\MVApair{s}{s'}}}             
\newcommand{\azx}[1]{\azimuth_{\scriptscriptstyle n,#1}^{\MVApair{s}{s'}}}               
\newcommand{\delayx}[1]{\widetilde{\tau}_{\scriptscriptstyle n,#1}^{\MVApair{s}{s'}}}    

\newcommand{\delay}[0]{\tau}        

\newcommand{\Ptemplate}[0]{\bm{P}_{\text{t}}}                       

\newcommand{\onematrix}[2]{\bm{1}_{\scriptscriptstyle{#1\times#2}}}        

\renewcommand{\etarot}[1]{\bm{\eta}^{\text{\tiny{o}}}_{\scriptscriptstyle#1}}
\newcommand{\etarotHat}[1]{\widehat{\bm{\eta}}^{\text{\tiny{o}}}_{\scriptscriptstyle#1}}
\newcommand{\toffsetHat}[1]{\widehat{\epsilon}_{\scriptscriptstyle#1}}    
\newcommand{\factor}[2]{f_{\scriptscriptstyle #1#2}}   
\renewcommand{\state}[1]{\bm{\theta}_{\scriptscriptstyle #1}}
\renewcommand{\observation}[2]{\bm{z}_{\scriptscriptstyle #1,#2}}   

\title{Spatiotemporal Synchronization of Distributed Arrays using Particle-Based Loopy Belief Propagation}

\twoauthors{Benjamin J. B. Deutschmann\sthanks{
The AMBIENT-6G project received funding from the EU Horizon Europe research and innovation program under grant agreement No~101192113.}}
{Graz University of Technology\\	
Graz, 8010 Austria}
{Peter Vouras%
}
{United States Department of Defense\\
Washington, DC, 20375 USA}
\usepackage[preprint]{spconf}

\copyrightnotice{CISA 2025}
\usepackage[
    type={CC},
    modifier={by},
    version={4.0},
]{doclicense}
\toappear{\doclicenseImage[imagewidth=5em]}

\newcommand{\shortPaperVersion}{true} 

\begin{document}
%
\maketitle
\begin{abstract}
Sensing and imaging with distributed radio infrastructures (e.g., distributed MIMO, wireless sensor networks, multistatic radar) rely on knowledge of the positions, orientations, and clock parameters of distributed apertures. We extend a particle-based loopy belief propagation (BP) algorithm to cooperatively synchronize distributed agents to anchors in space and time. Substituting marginalization over nuisance parameters with approximate but closed-form concentration, we derive an efficient estimator that bypasses the need for preliminary channel estimation and operates directly on noisy channel observations. 
Our algorithm demonstrates scalable, accurate spatiotemporal synchronization on simulated data.
\end{abstract}
\section{Introduction}\label{sec:intro}

\setlength{\abovedisplayskip}{3pt}
\setlength{\abovedisplayshortskip}{3pt}
\setlength{\belowdisplayskip}{3pt}
\setlength{\belowdisplayshortskip}{3pt}
\setlength{\textfloatsep}{5pt}  

Distributed antenna arrays exploit spatially separated nodes to achieve higher aperture gain, interference immunity, and improved spatial diversity for sensing or communications. 
However, such arrays require precise synchronization of time, phase, and frequency to ensure coherent summation of the signal carrier at the location of a target~\cite{Larsson24massiveSynchrony,Callebaut24Sync}.
This is especially challenging at millimeter-wave frequencies and GHz-wide modulation bandwidths, where clocks at different nodes must remain aligned despite drift.
Optical links can synchronize spatially separated clocks at femtosecond precision~\cite{Ye03optical} but are difficult to implement in dynamic scenarios due to strict pointing/tracking requirements and large form factors. 
Therefore, \gls{ota} wireless time synchronization techniques at microwave and millimeter-wave frequencies are of great interest~\cite{Mahmood18OTA,Larsson24massiveSynchrony}. 
%
\begin{figure}[t]%
    \centering%
    \includegraphics[width = \linewidth]{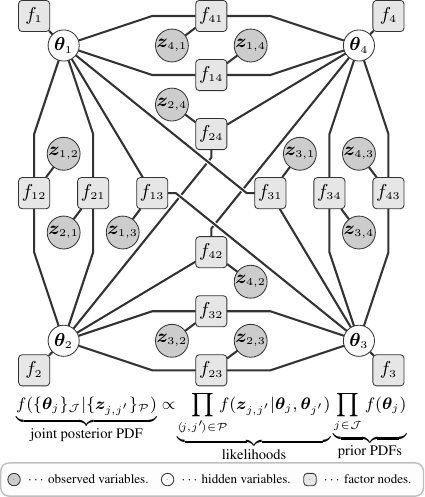}%
    \vspace{-0.2cm}\caption{Factor graph representing the joint posterior PDF. \\
    We abbreviate $\factor{j}{j'}:=f\big(\observation{j}{j'}|\state{j},\state{j'}\big)$ and $\factor{j}{}:=f\big(\state{j}\big)$.}%
    \label{fig:factor-graph}%
\end{figure}%
Prominent approaches are consensus-based time synchronization techniques~\cite{Rashid22ConsensusBP} because they scale with the size of the network and are robust to node failures. 
However, the accuracy of time synchronization is affected by message delays.
The message delays include propagation time and the transmission, media access, and reception process~\cite{Fang2022}.
In a dynamic scenario with moving platforms the random message delays may be orders of magnitude larger than the required precision and can cause synchronization algorithms to diverge.
\ifthenelse{\equal{\shortPaperVersion}{true}}%
{
}{%
The synchronization objective varies in strictness: determining only the correct ordering of events requires less accuracy than aligning clocks to a common absolute timeline. 
Relative synchronization translates time stamps between nodes, while absolute synchronization imposes a global definition of time.

There are fundamental limits inherent in synchronizing clocks over wireless networks~\cite{Freris2011}.  Determining all the unknown parameters including the skews and offsets of all the clocks and the message delays inherent in all the communication links is impossible.  The clock skews at each node as well as the round-trip delays between every pair of nodes can be determined correctly.  There are three notions of clock synchronization relevant to a wireless network.  First, is creating a correct sequence of chronological events across the entire network.  In this scenario, accurate estimation of exact time instants is not necessary.  Instead the ordering of events that occur at different nodes or the same node must be determined.  This model imposes the weakest requirements on synchronization.

Second, is the notion of relative synchronization.  The goal here is estimate the relative drift between a set of clocks in the network and use this information to translate time stamps from one clock to any other clock.  The third and hardest objective to achieve is absolute synchronization.  In this scenario, each node in the network operates an absolute clock and the goal is to bring all clock displays in agreement.  If successfully synchronized then the entire network operates under a global definition of time.  Note that absolute synchronization implies relative synchronization which subsequently implies a correct ordering of events in the network.  Further, if one network node is chosen as a reference, then relative synchronization can be used to attain absolute synchronization.
}%

{\slshape Contributions.} We build upon the efficient particle-based loopy \gls{bp} algorithm from~\cite{Meyer13particleBasedBP}, \cite[Sec.\,2.4]{Meyer15diss} where we extend the problem of inferring the positions and clock parameters of distributed apertures to \gls{3d} spatiotemporal synchronization, i.e., joint inference of the spatial states (positions and orientations) and the temporal states (clock parameters) of the distributed apertures.
While the original algorithm operates in a \textit{two-step} fashion (using intermediate channel parameter estimates such as delays), our algorithm works \textit{directly} on the noisy channel estimates between distributed apertures and does not require a dedicated channel estimator.
We leverage \textit{concentration} to eliminate nuisance parameters as an approximate but closed-form solution that avoids computationally involved marginalizations.

\section{System Model}\label{sec:system-model}
\newcommand{\setApertures}[0]{\mathcal{J}}      
\newcommand{\setAgents}[0]{\mathcal{M}}         
\newcommand{\setAnchors}[0]{\mathcal{A}}        
\newcommand{\setPairs}{\mathcal{P}}             
\newcommand{\setNeighbors}[1]{\mathcal{P}_{\scriptscriptstyle #1}}  
\newcommand{\Nchannel}[0]{N}                    
\newcommand{\naperture}[0]{j}                   
\newcommand{\nagent}[0]{m}                      
\newcommand{\nanchor}[0]{a}                     
\newcommand{\APpair}[2]{(\!#1,#2\!)}            
\newcommand{\stateHat}[1]{\hat{\bm{\theta}}_{\scriptscriptstyle #1}}
\renewcommand{\covarianceEstimate}[1]{\widehat{\bm{C}}_{\scriptscriptstyle#1}}                
\newcommand{\stateHatMMSE}[1]{\hat{\bm{\theta}}^{\text{\tiny MMSE}}_{\scriptscriptstyle #1}}
\newcommand{\amplitude}[2]{\alpha_{\scriptscriptstyle\!#1,#2}}
\renewcommand{\dictEntry}[2]{\bm{\psi}_{\scriptscriptstyle #1,#2}}  
\renewcommand{\noise}[2]{\bm{w}_{\scriptscriptstyle #1,#2}}           
\newcommand{\noiseVariance}[2]{\sigma^2_{\scriptscriptstyle #1,#2}} 
\renewcommand{\elx}[2]{\elevation_{\scriptscriptstyle #1,#2}^{\text{\tiny AoA}}}            
\renewcommand{\elxAoD}[2]{\elevation_{\scriptscriptstyle #1,#2}^{\text{\tiny AoD}}}      
\renewcommand{\azx}[2]{\azimuth_{\scriptscriptstyle #1,#2}^{\text{\tiny AoA}}}              
\renewcommand{\azxAoD}[2]{\azimuth_{\scriptscriptstyle #1,#2}^{\text{\tiny AoD}}}        
\renewcommand{\delayx}[2]{{\tau}_{\scriptscriptstyle #1,#2}}                      

\newcommand{\elAoAGeneral}[0]{\elevation^{\text{\tiny AoA}}}      
\newcommand{\azAoAGeneral}[0]{\azimuth^{\text{\tiny AoA}}}        
\newcommand{\elAoDGeneral}[0]{\elevation^{\text{\tiny AoD}}}      
\newcommand{\azAoDGeneral}[0]{\azimuth^{\text{\tiny AoD}}}        
\renewcommand{\rx}{r_{\scriptscriptstyle x}}
\renewcommand{\ry}{r_{\scriptscriptstyle y}}
\renewcommand{\rz}{r_{\scriptscriptstyle z}}
\renewcommand{\range}[2]{\bm{r}_{\scriptscriptstyle#1,#2}} 
\renewcommand{\ranget}[2]{\widetilde{\bm{r}}_{\scriptscriptstyle#1,#2}} 
\renewcommand{\rangepTX}[2]{\grave{\bm{r}}_{\scriptscriptstyle#1,#2}} 
\renewcommand{\rangep}[2]{\acute{\bm{r}}_{\scriptscriptstyle#1,#2}}     
\renewcommand{\rotM}[1]{\bm{M}_{\scriptscriptstyle#1}}                  
\newcommand{\Pj}[3]{\bm{P}_{\scriptscriptstyle #1}}                     
\newcommand{\amplitudeHat}[2]{\widehat{\alpha}_{\scriptscriptstyle\!#1,#2}}                     
\newcommand{\noiseVarianceHat}[2]{\widehat{\sigma^2}_{\scriptscriptstyle #1,#2}}                     
\renewcommand{\PiPerp}[2]{\bm{\Pi}_{\scriptscriptstyle {#1},{#2}}^{\perp}}  
We consider a set $\setApertures:= \{1 \dots J\} = \setAgents \cup \setAnchors$ of geometrically identical apertures comprising a set of agents $\setAgents$ with unknown spatiotemporal parameters and a set of anchors $\setAnchors$ with known spatiotemporal parameters
\begin{align}
    \state{j} :=\left[
        {\pos{j}}^\trp \ {\etarot{j}}^\trp \toffset{j}
    \right]^\trp 
    \quad \in \stateSpace
\end{align}
in \textit{global} coordinates where the state space $\stateSpace\!=\!\realsetone{7}$ comprises the position $\pos{j}\!\in\!\realsetone{3}$, orientation $\etarot{j}\!\in\!\realsetone{3}$, and clock time offset $\toffset{j}\in\realsetone{}$.
We assume that each aperture $\naperture'\!\in\!\setApertures$ transmits a pilot which the receiving apertures $\naperture\!\in\!\setApertures\!\setminus\!\naperture'$ use to compute noisy channel observations 
\begin{align}\label{eq:observation}
    \observation{j}{j'} := \amplitude{j}{j'} \, \dictEntry{j}{j'} (\state{j},\state{j'}) + \noise{j}{j'} \quad \in \complexsetone{\Nchannel}
\end{align}
taken in spatially and temporally uncorrelated noise $\noise{j}{j'}\!\sim\!\mathcal{CN}(\bm{0},\noiseVariance{j}{j'} \eye{\Nchannel})$ with amplitudes $\amplitude{j}{j'} \in \complexsetone{}$.
Let $\mathbb{S}^1\!:=\!\{x\!\in\!\mathbb{C}\!:\!|x|\!=\!1\}$ denote the $1$-sphere, i.e., the unit-circle, and let $\mathbb{T}^{\Nchannel}\!:=\!(\mathbb{S}^1)^\Nchannel$ denote the $\Nchannel$-torus, one particular steering vector $\dictEntry{j}{j'}(\state{j},\state{j'})\!\in\!\mathbb{T}^{\Nchannel}$ is from the spatiotemporal array manifold
\begin{align}\label{eq:dictEntry-channel}
    \bm{\psi} = 
        \bm{b}(\delay)  \otimes 
        \bm{a}(\elAoAGeneral\!,\!\azAoAGeneral) \otimes
        \bm{a}(\elAoDGeneral{}{}\!,\!\azAoDGeneral{}{})
\end{align}
modeled as the Kronecker product of the temporal manifold 
\begin{align}\label{eq:delay-manifold}
    \bm{b}(\delay) = \exp\big(\!-\!\mathrm{j} 2 \pi \bm{f} \delay\big) \quad \in \mathbb{T}^{\Nfrequency}
\end{align}
with the spatial manifolds in \gls{aoa} and \gls{aod}~\cite{richter2005estimation}
    $\bm{a}(\elevation,\azimuth) = \bm{a}_y(\elevation,\azimuth) \otimes \bm{a}_z(\elevation)$
each being itself a Kronecker product\footnote{Only URA layouts admit this Kronecker-factorization.} of the ``horizontal'' manifold
\begin{align}\label{eq:spatial-manifold-y}
    \bm{a}_y(\elevation,\azimuth) = \exp\Big(\mathrm{j} \frac{2 \pi}{\lambda} \bm{p}_y \sin(\elevation) \sin(\azimuth) \Big) \quad \in \mathbb{T}^{\Nantennasy}
\end{align}
and ``vertical'' manifold
\begin{align}\label{eq:spatial-manifold-z}
    \bm{a}_z(\elevation) = \exp\Big(\mathrm{j} \frac{2 \pi}{\lambda} \bm{p}_z \cos(\elevation) \Big) \quad \in \mathbb{T}^{\Nantennasz} \, 
\end{align}
for wavelength $\lambda$.
We use $\bm{f}\in\realsetone{\Nfrequency}$ to denote a vector of $\Nfrequency$ baseband frequencies, $\bm{p}_y\in \realsetone{\Nantennasy}$ a vector of $\Nantennasy$ horizontal sensor positions and $\bm{p}_z\in \realsetone{\Nantennasz}$ a vector of $\Nantennasz$ vertical sensor positions of a ``template'' aperture. 
The complete manifold in~\eqref{eq:dictEntry-channel} is of dimension $\Nchannel\triangleq\Nfrequency\Nantennasy^2\Nantennasz^2$.
We require each of the support vectors $\{\bm{f}, \bm{p}_y, \bm{p}_z\}$ to be symmetric around $0$. %
The steering vectors are elements of the manifolds parameterized by \textit{local} parameters in spherical aperture coordinates with
\begin{align}
    \delayx{j}{j'} = \frac{\lVert \rangep{j}{j'} \rVert}{\lightspeed} + (\toffset{j}-\toffset{j'})
\end{align}
denoting the perceived delay and $\lightspeed$ the propagation velocity,
\begin{align}
    \elevation(\bm{r}) = \arccos (\rz/ \lVert \bm{r} \rVert)
\end{align}
denoting elevation, and
\begin{align}
    \azimuth(\bm{r}) = \arctantwo (\ry,\rx)
\end{align}
denoting azimuth, for one arbitrary vector $\bm{r}=[r_x,r_y,r_z]^\trp$ 
in Cartesian aperture coordinates.

{\slshape Mapping from global to local parameters.} 
Let $\range{j}{j'} \!:=\! \pos{j'} \!-\! \pos{j}$ be the vector pointing from the phase center of aperture $j$ to aperture $j'$ in \textit{global} coordinates, we define the \gls{aoa} as $\{\elx{j}{j'},\azx{j}{j'}\}\!:=\!\{\elevation(\rangep{j}{j'}),\azimuth(\rangep{j}{j'})\}$
parameterized by the vector $\rangep{j}{j'} \!=\! \rotM{j}^{-1} \range{j}{j'}$ pointing from receiving aperture $j$ to transmitting aperture $j'$ in \textit{local} Cartesian coordinates of aperture $j$.
Likewise, we define the \gls{aod} as
$\{\elxAoD{j}{j'},\azxAoD{j}{j'}\}\!:=\{\elevation(\rangepTX{j}{j'}),\azimuth(\rangepTX{j}{j'})\}$
parameterized by the vector $\rangepTX{j}{j'} \!=\! \rotM{j'}^{-1} (-\range{j}{j'})$ pointing from transmitting aperture $j'$ to receiving aperture $j$ in \textit{local} Cartesian coordinates of aperture $j'$.
Let $\onematrix{M}{N}$ denote an $(M\times N)$-matrix of all ones, we model the \gls{ura} sensor layout 
\begin{align}
    \Ptemplate := 
    \begin{bmatrixs}
        \bm{0} \\
        \bm{p}_y^\trp \otimes \onematrix{1}{\Nantennasz} \\
        \onematrix{1}{\Nantennasy} \otimes \bm{p}_z^\trp \\
    \end{bmatrixs}
    \quad \in \realset{3}{\Nantennasy\Nantennasz}
\end{align}
of the template aperture in the $yz$-plane, symmetric about the origin $\origin$.
Abbreviating $\Nantennas:=\Nantennasy\Nantennasz$, the antenna positions of an actual \textit{physical} aperture are computed by shifting a \textit{template} aperture out of the origin $\origin$ to its center position $\pos{j}$ in global coordinates through~\cite{Deutschmann25OJSP}
\begin{align}\label{eq:PA-layout}
    \Pjss{j}{0}{0} =  \pos{j} \, \onematrix{1}{\Nantennas}\, + \rotM{j} \Ptemplate \quad \in \realset{3}{\Nantennas} .
\end{align}
The matrix $\rotM{j} \in SO(3)$ is a rotation matrix that defines the orientation of aperture $j$ in global coordinates.
It is from the special orthogonal group $SO(3)\!=\!\{\bm{M}\!\in\!\mathbb{R}^{3 \times 3}|\bm{M} \bm{M}^\trp =\bm{M}^\trp \bm{M}\!=\!\eye{3}, \det(\bm{M})\!=\!1 \}$. 
In this work, we use a parameterization $\rotM{j}(\etarot{j})$ with $\etarot{j}\!\in\!\realsetone{3}$ in Euler angles~\cite[p.\,86]{Kuipers99quaternions}, although a parameterization in unit quaternions~\cite[p.\,168]{Kuipers99quaternions} is likewise possible.
In the scenario (see Fig.\,\ref{fig:scenario}), all aperture orientations are chosen outside of the gimbal lock (e.g., the pitch angle $\pitch \neq \pm \nicefrac{\pi}{2}$).

\def\datapath{./figures}
\begin{figure}
\setlength{\figurewidth}{0.7\columnwidth}
    \centering
    \input{\datapath/scenario-CISA.tex}%
    \vspace{-3mm}%
    \caption{Scenario: Anchors $\setAnchors\!=\!\{1,4\}$ and agents $\setAgents\!=\!\{2,3\}$.
    }
    \label{fig:scenario}
\end{figure}

\section{Statistical Model}

{\slshape Deterministic Concentrated Likelihood.} 
We use a deterministic concentrated likelihood model assuming that the amplitudes $\amplitude{j}{j'}$ are \textit{deterministic} unknowns~\cite{KrimViberg96ASP}.
Per~\eqref{eq:observation}, the aperture-to-aperture 
channel observation is distributed as $\observation{j}{j'}|\state{j},\state{j'} \sim \mathcal{CN}(\amplitude{j}{j'} \dictEntry{j}{j'},\noiseVariance{j}{j'} \eye{\Nchannel})$.
Assuming independent observations between all $|\setPairs|\!=\!\frac{J!}{(J-2)!}\!=\!J(J\!-\!1)$ ordered aperture pairs $\APpair{j}{j'} \!\in\! \setPairs \!:=\! \big\{\APpair{j}{j'} \!\in\! \setApertures^2 | j\!\neq\! j'\big\}$, the joint likelihood factorizes as
\begin{equation}
    \begin{split}\label{eq:lhf-det}
    &f\big(\{\observation{j}{j'}\}_{\scriptscriptstyle \setPairs}|\{\state{j}\}_{\scriptscriptstyle \setApertures},\{\amplitude{j}{j'}, \noiseVariance{j}{j'}\}_{\scriptscriptstyle \setPairs}\big) 
    \\ 
    & = \!\!\prod_{\APpair{j}{j'}\in\setPairs}\!
    \frac{
        \exp \left( - \frac{1}{\noiseVariance{j}{j'}} \big\| \observation{j}{j'} - \dictEntry{j}{j'}(\state{j},\state{j'}) \amplitude{j}{j'} \big\|^2\right)
    }{
        \left(\pi \noiseVariance{j}{j'}\right)^{\Nchannel}
    } \,.
    \end{split}
\end{equation}
We compute the profile likelihood by concentrating w.r.t. the nuisance parameters $\{\amplitude{j}{j'}, \noiseVariance{j}{j'}\}_{\scriptscriptstyle \setPairs}$. 
That is, we compute \gls{ml} estimates conditional on the state pairs $\{\state{j},\state{j'}\}$ as~\cite{Stoica1995,Deutschmann24SPAWC,Deutschmann25OJSP}
\begin{align}
    \amplitudeHat{j}{j'}|\state{j},\!\state{j'} &= 
    \frac{1}{\Nchannel} ~ \dictEntry{j}{j'}^\herm \, \observation{j}{j'} ~,\label{eq:amplitudeHat}\\ 
    \noiseVarianceHat{j}{j'}|\state{j},\!\state{j'} &= \frac{1}{\Nchannel}  ~ \tr\Big( \PiPerp{j}{j'} \Rhate{j}{j'}\Big)~, \label{eq:noiseVarianceHat}
\end{align}
where $\PiPerp{j}{j'}(\state{j},\!\state{j'}\!)\!=\!\eye{\Nchannel}\!-\!\frac{\dictEntry{j}{j'}\dictEntry{j}{j'}^\herm}{\Nchannel}$ is the projector onto the noise subspace, and the 
matrix $\Rhate{j}{j'}\!:=\!\observation{j}{j'} \, \observation{j}{j'}^\herm$ 
is an \textit{empirical} rank-$1$ estimate of the 
noncentral second-moment matrix 
$\!\mathbb{E}\big(\observation{j}{j'}\observation{j}{j'}^\herm\big)$.
Although notationally brief,~\eqref{eq:noiseVarianceHat} can be implemented efficiently using \eqref{eq:noiseVarianceHatEfficient} in Appendix~\ref{app:implementation}.
Reinsertion of $\amplitudeHat{j}{j'}|\state{j},\!\state{j'}$ and $\noiseVarianceHat{j}{j'}|\state{j},\!\state{j'}$ in~\eqref{eq:lhf-det} yields the deterministic profile likelihood function\footnote{An efficient implementation uses $\observation{j}{j'}-\dictEntry{j}{j'}\amplitudeHat{j}{j}\triangleq \PiPerp{j}{j'}\observation{j}{j'}$.} for a single observation
\begin{equation}
    \begin{split}\label{eq:lhf-det-concentrated}
    &f\big(\observation{j}{j'}|\state{j},\state{j'}\big) =
    \frac{
        \exp \Big( - \frac{1}{\noiseVarianceHat{j}{j'}} \big\| \PiPerp{j}{j'}\observation{j}{j'}  \big\|^2\Big)
    }{
        \big(\pi \noiseVarianceHat{j}{j'}\big)^{\Nchannel}
    } \,.
    \end{split}
\end{equation}

{\slshape Factor Graph.} 
The joint posterior \gls{pdf} (up to a proportionality constant) of all aperture states conditional on all observations is represented by the factor graph~\cite{Loeliger04IntroFG} in Fig.\,\ref{fig:factor-graph} (for $J\!=\!4$) which---by exploiting the conditional independence structure that the graph makes explicit---can be shown to factorize as $f(\{\state{j}\}_{\scriptscriptstyle\setApertures}|\{\observation{j}{j'}\}_{\scriptscriptstyle\setPairs})%
\!\!\propto\!%
\!\prod_{\scriptscriptstyle\APpair{j}{j'}\in\setPairs} \!f(\observation{j}{j'}|\state{j},\state{j'})%
\!\prod_{\scriptscriptstyle j\in\setApertures}\!f(\state{j})$ into a product of the (concentrated) likelihoods in~\eqref{eq:lhf-det-concentrated} with prior \glspl{pdf} $f(\state{j})$ of the states of all apertures $j$. 
\ifthenelse{\equal{\shortPaperVersion}{true}}%
{
}{%
The prior \glspl{pdf} are used to incorporate beliefs about the states $\{\state{j}\}_{\setApertures}$ before observations are made, where they are practical to restrict the support of the parameter space to physically plausible solutions.
}%
We aim to compute the \gls{mmse} state estimate of aperture $j$ as the expectation under its marginal posterior \gls{pdf}: 
\begin{align}\label{eq:stateHatMMSE}
    \stateHatMMSE{j} = \mathbb{E}\big(\state{j}|\{\observation{j}{j'}\}_{\scriptscriptstyle\setPairs} \big)\,.
\end{align}
To obtain the marginal \glspl{pdf} $f(\state{j}|\{\observation{j}{j'}\}_{\scriptscriptstyle\setPairs})$, the \gls{spa}~\cite[Sec.\,8.4.4]{Bishop} would result in \textit{exact} inference in tree-structured factor graphs. 
However, problems of the class of cooperative localization result in graphs with loops where \textit{approximate} inference~\cite[Sec.\,V]{Kschischang01factorGraphs} is possible by iteratively applying the \gls{spa} resulting in loopy \gls{bp}~\cite{Frey97loopyBP}.%
\newcommand{\observationBar}[1]{\overline{\bm{z}}_{\scriptscriptstyle #1}}%
\newcommand{\stateBar}[1]{\overline{\bm{\theta}}}%
\newcommand{\stateTilde}[1]{\widetilde{\bm{\theta}}_{\scriptscriptstyle #1}}%
\newcommand{\belief}[1]{b^{\scriptscriptstyle (#1)}}
\newcommand{\beliefHat}[1]{\hat{b}^{\scriptscriptstyle (#1)}}
\newcommand{\beliefTilde}[1]{\tilde{b}^{\scriptscriptstyle (#1)}}
\newcommand{\particleBar}[2]{\overline{\bm{\theta}}_{\scriptscriptstyle }^{\scriptscriptstyle(#2)}}     
\newcommand{\particleTilde}[2]{\widetilde{\bm{\theta}}_{\scriptscriptstyle #1}^{\scriptscriptstyle(#2)}}     

\newcommand{\weightBar}[2]{\overline{w}_{\scriptscriptstyle #1}^{{\scriptscriptstyle(#2)}}}          
\newcommand{\weightTilde}[2]{\widetilde{w}_{\scriptscriptstyle #1}^{{\scriptscriptstyle(#2)}}}          
\newcommand{\normalizationConstant}[1]{{c}_{\scriptscriptstyle j, #1}}      %
\newcommand{\normalizationConstantBar}[1]{\overline{c}_{\scriptscriptstyle #1}}      %

{\slshape Loopy BP.} Let $\setNeighbors{j}\!:=\!\{ (\grave{\iota},\acute{\iota}) \!\in\! \setApertures^2 | (\grave{\iota} \!=\!j \lor  \acute{\iota} \!=\! j) \land \grave{\iota} \!\neq\! \acute{\iota} \} \!\subset\! \setPairs$ define the set of aperture pairs that involve aperture $j$.
Further, let $\observationBar{}\!:=\![\observation{j}{j'}]_{\scriptscriptstyle \APpair{j}{j'} \in \setPairs}$ 
and $\stateBar{j}\!:=\![\state{j}]_{\scriptscriptstyle j\in\setApertures}$, $\stateTilde{j}\!:=\![\state{j'}]_{\scriptscriptstyle j'\in\setApertures\setminus j}$ denote vectors of stacked~\cite{Meyer15diss} observations and states, respectively.
We devise iterative \gls{bp} message passing~\cite{Wymeersch09coopLocalization} where in each message passing iteration $p\!\in\!\{1\dots P\}$ we compute the beliefs~\cite{Meyer15diss}
\begin{align}\label{eq:belief}
    \belief{p}\!(\state{j}) \!&\propto \!\!\int \!\! \belief{p}(\stateBar{}) \diff \stateTilde{j}
    \\
    &= \! \!f(\state{j}) \!\!\int 
    \!\!
    \!\prod_{\scriptscriptstyle (\grave{\iota},\acute{\iota})\in\setPairs} \!\!\!f(\observation{\grave{\iota}}{\acute{\iota}}|\state{\grave{\iota}},\state{\acute{\iota}}) 
    \!\prod_{\scriptscriptstyle j'\!\in\setApertures\setminus j}\!\!\!\belief{p\!-\!1}\!(\state{j'\!}) \diff \stateTilde{j} \nonumber \\
    &\propto \! \!f(\state{j}) \!\!\int 
    \!\!
    \!\prod_{\scriptscriptstyle (\grave{\iota},\acute{\iota})\in\setNeighbors{j}} \!\!\!f(\observation{\grave{\iota}}{\acute{\iota}}|\state{\grave{\iota}},\state{\acute{\iota}}) 
    \!\prod_{\scriptscriptstyle j'\!\in\setApertures\setminus j}\!\!\!\belief{p\!-\!1}\!(\state{j'\!}) \diff \stateTilde{j} \nonumber 
\end{align}
which are approximations of the marginal posterior \glspl{pdf} $f(\state{j}|\observationBar{})$.
Note that the marginalization integral reduces the likelihood factor multiplication over the set of \textit{all} aperture pairs $\setPairs$ to a multiplication over the set $\setNeighbors{j}\subset \setPairs$ of \textit{neighboring} likelihood factors connected to aperture $j$, thereby exploiting the conditional independence structure of the graph.
According to the rules of the \gls{spa}, the marginal posterior \gls{pdf} $f(\state{j}|\observationBar{})$ of aperture $j$ computes as the product of all incoming messages, one of which is the prior $f(\state{j})$ of node $j$.
In a graph such as in Fig.\,\ref{fig:factor-graph}, the other messages compute as the marginalization integrals of the product of the marginal posterior \glspl{pdf} of all other nodes $j'\in\setApertures\setminus j$ with the likelihood factors connecting them to node $j$ (i.e., $ \prod_{(\grave{\iota},\acute{\iota}) \in\setNeighbors{j}}\factor{\grave{\iota}}{\acute{\iota}}$).
Since the graph has loops, the marginal posterior \glspl{pdf}, approximated by the beliefs $\belief{p}\!(\state{j})$, of all nodes $j$ 
depend on the marginal posterior \glspl{pdf} of all other nodes and are hence updated iteratively using loopy \gls{bp}.
\section{Particle-Based Implementation}\label{sec:implementation}%
Direct computation of marginalization integrals can be computationally involved or prohibitive.
We resort to a particle-based implementation based on~\cite{Meyer13particleBasedBP,Wielandner21PosOrient,Wielandner9Dcooperative,Leitinger24mpSLAMmapFusion} where we approximate the \textit{joint} belief as $\beliefHat{p}(\stateBar{})\!:=\!\sum_{\scriptscriptstyle i=1}^{\Nparticles}\!\weightBar{p|p}{i} \delta\big(\stateBar{} - \particleBar{p}{i}\big)\!\!\approx\!\!f(\stateBar{j}|\observationBar{})$ using a random measure $\{\weightBar{p|p}{i},\particleBar{j}{i}\}_{\scriptscriptstyle i=1}^{\scriptscriptstyle\Nparticles}$ of $\Nparticles$ particles $\particleBar{j}{i}$ with weights $\weightBar{p|p}{i}$ s.t. $\sum_{\scriptscriptstyle i=1}^{\Nparticles}\weightBar{p|p}{i} \!\triangleq\!1$, ensured by the normalization constant $\normalizationConstantBar{p}$. 
Using $\weightBar{p|p}{i}\!=\!\normalizationConstantBar{p}^{\scriptscriptstyle-1}\!\prod_{\scriptscriptstyle (\grave{\iota},\acute{\iota})\in\setPairs}\!f(\observation{\grave{\iota}}{\acute{\iota}}|\state{\grave{\iota}},\state{\acute{\iota}}) \weightBar{p|p-1}{i}$, the particle-based approximation of the beliefs in~\eqref{eq:belief} is 
\begin{align}\label{eq:beliefHat}
    \beliefHat{p}\!&(\state{j}) \!\propto \!\!\int \!\! \beliefHat{p}(\stateBar{}) \diff \stateTilde{j} = \!\!\int  \sum_{\scriptscriptstyle i=1}^{\Nparticles}\!\weightBar{p|p}{i} \delta\big(\stateBar{} \!-\! \particleBar{p}{i}\big) \diff \stateTilde{j}\\[-8pt]
    &\propto \! 
    \!\!\int 
    \!\!
    \sum_{\scriptscriptstyle i=1}^{\Nparticles} \normalizationConstantBar{p}^{\scriptscriptstyle-1}\!\!\!\!
    \!\prod_{\scriptscriptstyle (\grave{\iota},\acute{\iota})\in\setNeighbors{j}} \!\!\!f(\observation{\grave{\iota}}{\acute{\iota}}|\state{\grave{\iota}},\state{\acute{\iota}}) 
        \weightBar{p|p-1}{i} \delta\big(\stateBar{} \!-\! \particleBar{p}{i}\big) 
    \diff \stateTilde{j}\nonumber 
    \\[-5pt]
    &\propto \! 
    \!\!
    \sum_{\scriptscriptstyle i=1}^{\Nparticles} 
    \overbrace{
        \normalizationConstant{p}^{\scriptscriptstyle-1}\!\!\!
        \!\prod_{\scriptscriptstyle (\grave{\iota},\acute{\iota})\in\setNeighbors{j}} \!\!\!f(\observation{\grave{\iota}}{\acute{\iota}}|\state{\grave{\iota}}^{\scriptscriptstyle(i)},\state{\acute{\iota}}^{\scriptscriptstyle(i)}) 
    \weightBar{p|p-1}{i}  
    }^{=:\weight{j,p}{i}}
    \delta\big(\state{j} \!-\! \particle{j}{i}\big)
    \nonumber 
\end{align}
with $\normalizationConstant{p}\!=\!\sum_{\scriptscriptstyle i=1}^{\scriptscriptstyle\Nparticles}\prod_{\scriptscriptstyle (\grave{\iota},\acute{\iota})\in\setNeighbors{j}} \!f(\observation{\grave{\iota}}{\acute{\iota}}|\state{\grave{\iota}}^{\scriptscriptstyle(i)},\state{\acute{\iota}}^{\scriptscriptstyle(i)}) \weightBar{p|p-1}{i}$.
At $p\!=\!0$, we initialize particles $\particle{j}{i} \!\!\!\sim \mathcal{U}(\state{\text{\tiny min}}^{\scriptscriptstyle(j)},\state{\text{\tiny max}}^{\scriptscriptstyle(j)}) \, \forall j\!\in\!\setApertures$ drawing from a uniform distribution, which absorbed the prior factor $f(\state{j})$ in~\eqref{eq:belief} and causes $\weightBar{0|0}{i}\!=\!\nicefrac{1}{\Nparticles} \, \forall i\!\in\{1\hdots\Nparticles\}$.
We estimate the state $\state{j}$ by approximating the \gls{mmse} estimate in~\eqref{eq:stateHatMMSE}
\begin{align}\label{eq:state-estimate}
    \stateHat{j}^{\scriptscriptstyle (p)} = \int_{\stateSpace}\!\state{j} \,\beliefHat{p}(\state{j}) \,  \diff \state{j}= \sum\nolimits_{i=1}^{\Nparticles}  \particle{j}{i} \weight{j,p}{i} ~ ,
\end{align}
and approximate the state covariance matrix as
\begin{align}\label{eq:empiricalCovariance}
    \covarianceEstimate{j}^{\scriptscriptstyle (p)} =  \sum\nolimits_{i=1}^{\Nparticles}
    \left(\particle{j}{i} - \stateHat{j}^{\scriptscriptstyle (p)}\right)\left(\particle{j}{i} - \stateHat{j}^{\scriptscriptstyle (p)}\right)^{\!\trp} \weight{j,p}{i} \, .
\end{align}
We implement loopy \gls{bp} using Alg.\,\ref{alg:loopyBP}. 
In line~\ref{line:resampling}, we employ systematic resampling~\cite[Alg.\,2]{Arulampalam02PFtutorial} 
which reduces particle degeneracy and implies equal weights after resampling (see line~\ref{line:resampled-weights}).
After resampling, each particle is convolved (see line~\ref{line:kernelConvolution}) with a Gaussian regularization kernel $K(\state{j})$ using the 
covariance matrix $\covarianceEstimate{j}^{\scriptscriptstyle (p)}$ and scaled by the optimal kernel bandwidth $h_{\text{\tiny opt}}$~\cite[p.\,253]{Musso2001PF}, where $\bm{\nu}_i\!\sim\!\mathcal{N}(\bm{0},\eye{ })$, 
to counteract particle impoverishment.
Following~\cite[Sec.\,2.4.1]{Meyer15diss}, importance sampling from the proposal distribution $\belief{p\!-\!1}(\stateBar{j})$ is implemented implicitly by stacking particles. 
\begin{algorithm}[t] 
	\LinesNumbered		
	\caption{Regularized Particle-Based Loopy BP}\label{alg:loopyBP}
	\SetKwInOut{Input}{Input}\SetKwInOut{Output}{Output}
	\Input{$\Nparticles,\{\state{\text{\tiny min}}^{\scriptscriptstyle(j)},\state{\text{\tiny max}}^{\scriptscriptstyle(j)}\}_{j\in\setApertures},\{\observation{j}{j'}\}_{(j,j')\in\setPairs}$} 
	\Output{$\{\stateHat{j}^{\scriptscriptstyle (P)}\}_{j\in\setAgents}$}	
	$\{\particle{j}{i} \!\!\!\sim \mathcal{U}(\state{\text{\tiny min}}^{\scriptscriptstyle(j)},\state{\text{\tiny max}}^{\scriptscriptstyle(j)})\}_{j\in\setAgents}$,  $\weightBar{0|0}{i}\!\!\gets\!\nicefrac{1}{\Nparticles} \, \forall i\!\in\! \{1 \dots \Nparticles\}$\; 
	  \For{$p \gets 1$ \KwTo $P$ \KwBy $1$}{ 
            \For{$\APpair{j}{j'} \in \setPairs$\label{line:pairloop}}{ 
	  	\For{$i \gets 1$ \KwTo $\Nparticles$ \KwBy $1$\label{line:parfor}}{ 
            $\weightTilde{j,j'}{i} \gets f(\observation{j}{j'}|\state{j}^{\scriptscriptstyle(i)}\!\!,\state{j'}^{\scriptscriptstyle(i)})$\; 
		}
            $\big\{\weight{j,j'}{i} \big\}_{\scriptscriptstyle i=1}^{\scriptscriptstyle\Nparticles} \gets \big\{ \weightTilde{j,j'}{i}\big/\big(\sum_{\scriptscriptstyle i=1}^{\scriptscriptstyle\Nparticles} \weightTilde{j,j'}{i}\big) \big\}_{\scriptscriptstyle i=1}^{\scriptscriptstyle\Nparticles}$\;
            }
        \For{$j \gets 1$ \KwTo $J$ \KwBy $1$}{ 
        $\big\{\weightBar{p|p-1}{i}\big\}_{\scriptscriptstyle i=1}^{\scriptscriptstyle\Nparticles}\!\! \gets\! \big\{\frac{f(\particle{j}{i})}{\beliefTilde{p-1}(\particle{j}{i})}\weightBar{p-1|p-1}{i}\big\}_{\scriptscriptstyle i=1}^{\scriptscriptstyle\Nparticles}$;\hfill \textcolor{gray}{//\,\textsuperscript{${(\ast)}$} \hspace{-4mm}}\label{line:importance-weight-compensation} \\
        $\big\{\weightTilde{j,p}{i}\big\}_{\scriptscriptstyle i=1}^{\scriptscriptstyle\Nparticles} \gets \big\{\weightBar{p|p-1}{i}~\prod_{\scriptscriptstyle\APpair{j}{j'}\in\setNeighbors{j}} \weight{j,j'}{i} \big\}_{\scriptscriptstyle i=1}^{\scriptscriptstyle\Nparticles} $\;
        $\big\{\weight{j,p}{i} \big\}_{\scriptscriptstyle i=1}^{\scriptscriptstyle\Nparticles} \gets \big\{ \weightTilde{j,p}{i}\big/\big(\sum_{\scriptscriptstyle i=1}^{\scriptscriptstyle\Nparticles} \weightTilde{j,p}{i}\big) \big\}_{\scriptscriptstyle i=1}^{\scriptscriptstyle\Nparticles}$\;
	$\stateHat{j}^{\scriptscriptstyle (p)}  \gets \sum_{i=1}^{{\Nparticles}}  \particle{j}{i} \weight{j,p}{i}$; \hfill \textcolor{gray}{//see \eqref{eq:state-estimate}\hspace{-4mm}}\\
        $\covarianceEstimate{j}^{\scriptscriptstyle (p)}  \gets \sum_{i=1}^{{\Nparticles}} \big(\particle{j}{i} - \stateHat{j}\big)\!\big(\particle{j}{i} - \stateHat{j}\big)^\trp \!\weight{j,p}{i}$\;
        $\{\particle{j}{i}\}_{\scriptscriptstyle i=1}^{{\scriptscriptstyle\Nparticles}}\!\gets\!\texttt{\small resample}(\{\particle{j}{i},\weight{j,p}{i}\}_{\scriptscriptstyle i=1}^{{\scriptscriptstyle\Nparticles}})$; \hfill \textcolor{gray}{//\cite
        {Arulampalam02PFtutorial}\hspace{-4mm}}\label{line:resampling} \\
        $\{\weightBar{p|p}{i}\}_{\scriptscriptstyle i=1}^{{\scriptscriptstyle\Nparticles}} \gets 1/{\Nparticles}$;\hfill \textcolor{gray}{//due to resampling\hspace{-4mm}}\label{line:resampled-weights}\\
        $\bm{L}_j \gets \texttt{\small cholesky}(\covarianceEstimate{j}^{\scriptscriptstyle (p)})$; \hfill \textcolor{gray}{//s.t. $\bm{L}_j \bm{L}_j^\trp=\covarianceEstimate{j}^{\scriptscriptstyle (p)} $\hspace{-4mm}}\\
        \For{$i \gets 1$ \KwTo ${\Nparticles}$ \KwBy $1$}{ 
            $\bm{\nu}_i \sim \mathcal{N}(\bm{0},\eye{ })$\;
            $\particle{j}{i} \gets \particle{j}{i} + h_{\text{\tiny opt}} \bm{L}_j \bm{\nu}_i$\label{line:kernelConvolution}\;
        }%
        }%
        }%
\vspace{-0.1cm}     
\algorithmfootnote{\textsuperscript{${(\ast)}$}\footnotesize\,Computing~\eqref{eq:beliefHat}, drawing $\{\particle{j}{i}\}_{\scriptscriptstyle i=1}^{\scriptscriptstyle\Nparticles}$ from ${f}(\state{j})$ is not appropriate if $f(\state{j})$ is not informative~\cite{Meyer2015DistLocExtended}. 
Drawing from $\beliefHat{p-1}(\state{j})$ instead introduces the importance ratio in line~\ref{line:importance-weight-compensation}, but its support is lost after resampling.
Kernel density estimation can circumvent this problem, but loses scalability~\cite{Meyer15diss}.
We use a unimodal Gaussian surrogate approximation $\beliefTilde{p-1}(\particle{j}{i}):=\mathcal{N}(\particle{j}{i};\stateHat{j}^{\scriptscriptstyle (p-1)},\covarianceEstimate{j}^{\scriptscriptstyle (p-1)})$ sustaining scalability, yet avoiding degeneracy.
\vspace{0.3cm}
}
\end{algorithm}
%
%
%
\section{Results}\label{sec:results}
\newcommand{\MCruns}{10000}
We assume to observe $\Nfrequency\!=\!10$ frequency bins within a bandwidth of $B\!=\!\SI{500}{\mega\hertz}$ centered around $\fc\!=\!\SI{6.175}{\giga\hertz}$ which matches with the \gls{nr} band 
n102~\cite{ETSI_TS_138_101}, 
also used for Wi-Fi\,6E defined in IEEE\,Std.\,802.11ax~\cite{IEEE_802-11ax}. 
The channel input~\gls{snr}~\cite{Deutschmann23ICC} is $\SI{10}{\dB}$ and the array spacing $d\!=\!\frac{\lambda}{2}$.
We simulate the scenario in Fig.\,\ref{fig:scenario} with $J\!=\!4$ apertures equipped with $(4\!\times\!4)$-\glspl{ura} with anchors $\setAnchors\!=\!\{1,4\}$ (initialized as $\particle{j}{i}\!\!=\!\state{j} \,\forall j\!\in\! \setAnchors, i\!\in\!\{1 \hdots \Nparticles\}$) and agents $\setAgents\!=\!\{2,3\}$ (initialized as $\particle{j}{i} \!\!\!\sim \mathcal{U}(\state{\text{\tiny min}}^{\scriptscriptstyle },\state{\text{\tiny max}}^{\scriptscriptstyle })\,\forall j\!\in\!\setAgents, i\!\in\!\{1 \hdots \Nparticles\}$). 
We choose $\Nparticles\!=\!\num{10000}$ particles and perform \num{\MCruns} estimation runs of Alg.\,\ref{alg:loopyBP} with random initializations.
With $\posEstimate{j}\!:=\![\stateHat{j}]_{\scriptscriptstyle1:3}$ and $\etarotHat{j}\!:=\![\stateHat{j}]_{\scriptscriptstyle4:6}$, Fig.\,\ref{fig:CDFs-combined} shows the cumulative frequency of the position errors 
$\lVert \posEstimate{j}\!-\!\pos{j} \rVert$ 
and the orientation errors 
$\lVert \etarotHat{j}\!-\!\etarot{j} \rVert$, after removing \SI{4.7}{\percent} of diverged runs. 
Fig.\,\ref{fig:RMSE-vs-iterations} shows the convergence behavior of the algorithm versus message passing iterations $p$.
\newcommand{\markRepeat}{15}
\newcommand{\markSize}{1}
\tikzexternaldisable
{
\renewcommand{\trajectoryLW}{0.7pt}     
\begin{figure}[tbp]
  \centering
    \vspace{-0.1cm}     
    \centering
    \setlength{\figurewidth}{0.85\columnwidth}
    \setlength{\figureheight}{0.175\columnwidth}
    \input{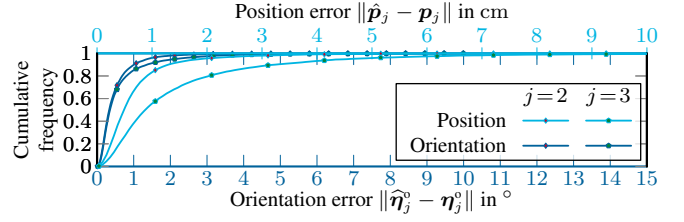}
    \vspace{-0.5cm}
  \caption{Cumulative freq. of the error of \gls{mmse} estimates. 
  }\label{fig:CDFs-combined}
\end{figure}
}
\tikzexternalenable
\tikzexternaldisable
\begin{figure}[t]
\raggedleft
\setlength{\figurewidth}{0.94\columnwidth}
\setlength{\figureheight}{0.15\columnwidth}
%
%
%

\pgfplotsset{every axis/.append style={
  label style={font=\footnotesize},
  legend style={font=\footnotesize},
  tick label style={font=\footnotesize},
}}

\definecolor{IEEEblue}{RGB}{0 98 155}
\definecolor{IEEEorange}{RGB}{225 163 0}

\begin{tikzpicture}[spy using outlines={circle, magnification=3.25, size=1.0cm, connect spies}]

\begin{axis}[%
name=boundary,  
width=0.951\figurewidth,
height=\figureheight,
at={(0\figurewidth,0\figurewidth)},
scale only axis,
tick align=inside,
axis line style = thick,	    
xmin=1,
xmax=50,
ylabel style = {yshift=-0.3mm},
ymode=log,     
ymin=0.75,
ymax=180,
ylabel={$\lVert \posEstimate{j} - \pos{j} \rVert$},
ytick=      { 1, 3,   10,  30, 180},
yticklabels={ 1, 3,   10,  30, 180},
xticklabel=\empty,                      
grid=none,          
line cap = round,
line join = round,
xmajorgrids,
ymajorgrids,
zmajorgrids,
major grid style={thin,loosely dotted}, 
legend style={legend cell align=left, align=left, draw=white!5!black,fill=white, fill opacity=0.90,font=\scriptsize} 
]

\addplot [color=AP2color, line width=\trajectoryLW, line join=round, line cap=round, 
]
  table[row sep=crcr]{%
1	180.788125055078\\
2	179.628479320984\\
3	175.832109170148\\
4	166.409192511259\\
5	150.494558595378\\
6	129.845151326406\\
7	105.855957764366\\
8	81.3141415999288\\
9	59.3636002859032\\
10	41.1374937615144\\
11	27.5412262394929\\
12	17.7095753511228\\
13	11.3767857015335\\
14	7.67173108813421\\
15	5.56252180311403\\
16	4.42739759094182\\
17	3.71012137454952\\
18	3.29030056233848\\
19	3.06936345268081\\
20	2.8898895413255\\
21	2.76028589387788\\
22	2.59881475646812\\
23	2.44302085080263\\
24	2.31021472663074\\
25	2.18719758540234\\
26	2.0723101126725\\
27	1.96418490694257\\
28	1.86316988135881\\
29	1.76850506436113\\
30	1.67959734971018\\
31	1.5962257079725\\
32	1.51979926817863\\
33	1.44792283155116\\
34	1.37992672868746\\
35	1.31545175112219\\
36	1.25701603280195\\
37	1.20353394685359\\
38	1.15129206377796\\
39	1.1026467601053\\
40	1.05709885709618\\
41	1.01601739602247\\
42	0.977788278794811\\
43	0.939487132662743\\
44	0.906761431934934\\
45	0.87274283074397\\
46	0.843219334703158\\
47	0.814124876574452\\
48	0.78728747271227\\
49	0.761729923896395\\
50	0.738272703323476\\
51	0.715548797204512\\
};\label{pgf:AP2}

\addplot [color=AP3color, line width=\trajectoryLW, line join=round, line cap=round, 
]
  table[row sep=crcr]{%
1	140.657156636416\\
2	140.354656855905\\
3	139.964222462629\\
4	139.340909074971\\
5	138.036347959742\\
6	135.031907958352\\
7	129.81211643024\\
8	121.584617855985\\
9	109.772003683218\\
10	94.596529008189\\
11	77.2998789393933\\
12	60.0284672728694\\
13	43.6841757328752\\
14	30.7122296586342\\
15	20.963466235999\\
16	14.1208643566974\\
17	9.90099751747351\\
18	7.90989218984633\\
19	7.1503399612969\\
20	6.77600655701065\\
21	6.35557269429116\\
22	5.81545554979006\\
23	5.5376017317913\\
24	5.12206685288596\\
25	4.76713545192658\\
26	4.48607796927829\\
27	4.2287008515881\\
28	3.9969226459308\\
29	3.78545160958538\\
30	3.58742530194966\\
31	3.40533732239729\\
32	3.23268562771849\\
33	3.0755111463193\\
34	2.93260634795826\\
35	2.80047716239358\\
36	2.67305745813596\\
37	2.55406226188001\\
38	2.44712479445863\\
39	2.34199060664536\\
40	2.24268508110098\\
41	2.14874097576534\\
42	2.06121244679836\\
43	1.98064761643189\\
44	1.90023685571556\\
45	1.82616072357555\\
46	1.75485161902842\\
47	1.69256930201034\\
48	1.63106270030979\\
49	1.57164546153561\\
50	1.51618059991101\\
51	1.4660502627151\\
};\label{pgf:AP3}

\end{axis}

\node[draw,fill=white,inner sep=1pt,below=-0.0em,xshift=-0.0mm,line cap = round, line join = round, line width = 0.5pt]
at (boundary.north) {\footnotesize \color{black}\,a) Position (\SI{}{\centi\metre})\,};

\newcommand{\columnSpace}{-1.15mm}   

\node[left, align=right ,font={\footnotesize}, draw,fill=white,inner sep=0.4pt,below left=-0.0em,yshift=-0.0mm,xshift=-0.1mm,line cap = round, line join = round, line width = 0.5pt] 
at (boundary.north east) {\scriptsize 
    \begin{tabular}{c l l l}
    \hspace{\columnSpace}\ref{pgf:AP2} $j\!=\!2$ \\
    \hspace{\columnSpace}\ref{pgf:AP3} $j\!=\!3$ 
    \end{tabular}};

\end{tikzpicture}%
%
\vspace{-1mm}%
%
%
%

\pgfplotsset{every axis/.append style={
  label style={font=\footnotesize},
  legend style={font=\footnotesize},
  tick label style={font=\footnotesize},
}}

\definecolor{IEEEblue}{RGB}{0 98 155}
\definecolor{IEEEorange}{RGB}{225 163 0}

\begin{tikzpicture}[spy using outlines={circle, magnification=3.25, size=1.0cm, connect spies}]

\begin{axis}[%
name=boundary,  
width=0.951\figurewidth,
height=\figureheight,
at={(0\figurewidth,0\figurewidth)},
scale only axis,
tick align=inside,
axis line style = thick,	    
xmin=1,
xmax=50,
ylabel style = {yshift=-0.3mm},
ymode=log,      
ymin=0.5,
ymax=179,
ylabel={$\lVert \etarotHat{j} - \etarot{j} \rVert$},
ytick=      {  1, 3,  10,  30,   100, 160},
yticklabels={  1, 3,  10,  30,   100},
xticklabel=\empty,                      
grid=none,          
line cap = round,
line join = round,
xmajorgrids,
ymajorgrids,
zmajorgrids,
major grid style={thin,loosely dotted}, 
legend style={legend cell align=left, align=left, draw=white!5!black,fill=white, fill opacity=0.90,font=\scriptsize} 
]

\addplot [color=AP2color, line width=\trajectoryLW, line join=round, line cap=round, 
]
  table[row sep=crcr]{%
1	101.92428367566\\
2	101.249492626313\\
3	99.1043048600717\\
4	93.8199557029614\\
5	84.9311073375023\\
6	73.4078683240761\\
7	60.0973486249075\\
8	46.4243883142835\\
9	34.0488528817577\\
10	23.7196102304928\\
11	15.8934804776586\\
12	10.1709958225207\\
13	6.47165904382179\\
14	4.284017140532\\
15	3.03180086517521\\
16	2.37045185822029\\
17	1.99390549694963\\
18	1.79426229011341\\
19	1.7120073129836\\
20	1.64683126962134\\
21	1.59927247184809\\
22	1.53846103550998\\
23	1.46849871485742\\
24	1.41225724709248\\
25	1.35863787957625\\
26	1.30562558708067\\
27	1.25578306516457\\
28	1.2066310177317\\
29	1.15952050044715\\
30	1.11388953666244\\
31	1.07015511189259\\
32	1.02853412904088\\
33	0.98884893798572\\
34	0.950301221837382\\
35	0.912792136096489\\
36	0.878659154219354\\
37	0.845116332489125\\
38	0.813048415686763\\
39	0.782719836333062\\
40	0.752743387988369\\
41	0.725215294974447\\
42	0.698971434424553\\
43	0.672480227893519\\
44	0.648490807661318\\
45	0.624677627122719\\
46	0.602988418233534\\
47	0.581543214988561\\
48	0.561477452569508\\
49	0.542230227407877\\
50	0.52385202626858\\
51	0.506184076330298\\
};

\addplot [color=AP3color, line width=\trajectoryLW, line join=round, line cap=round, 
]
  table[row sep=crcr]{%
1	160.609983153012\\
2	160.583689551283\\
3	160.516277016532\\
4	160.242950588573\\
5	159.246102073608\\
6	156.352745102362\\
7	150.935796614767\\
8	142.071492205501\\
9	128.9306727642\\
10	111.64106034785\\
11	91.6494996175747\\
12	71.2369888167552\\
13	51.1181272838732\\
14	34.7006822927777\\
15	21.8049384892927\\
16	12.1662549204305\\
17	5.86296531318885\\
18	3.16338264538416\\
19	2.63637473621492\\
20	2.52265577824357\\
21	2.28416463448125\\
22	2.11237992498228\\
23	1.9925574283252\\
24	1.83492303709834\\
25	1.72175182978631\\
26	1.63985520601595\\
27	1.5646567752643\\
28	1.49637051033819\\
29	1.43178101506992\\
30	1.37108972419268\\
31	1.31447031334207\\
32	1.25887877173805\\
33	1.20715096639321\\
34	1.16027051523644\\
35	1.1165294391379\\
36	1.07351730156378\\
37	1.03291555675568\\
38	0.996108197962145\\
39	0.959889531656639\\
40	0.92454677551755\\
41	0.891984346731149\\
42	0.860445347999082\\
43	0.830877363919617\\
44	0.801891548366433\\
45	0.774074306141813\\
46	0.74757093302782\\
47	0.723609443146967\\
48	0.700422743303029\\
49	0.677494819116334\\
50	0.656692889069346\\
51	0.638673577833531\\
};
\end{axis}

\node[draw,fill=white,inner sep=1pt,below=-0.0em,xshift=-0.0mm,line cap = round, line join = round, line width = 0.5pt]
at (boundary.north) {\footnotesize \color{black}\,b) Orientation (\SI{}{\degree})\,};

\end{tikzpicture}%
\vspace{-1mm}
%
%
%

\pgfplotsset{every axis/.append style={
  label style={font=\footnotesize},
  legend style={font=\footnotesize},
  tick label style={font=\footnotesize},
}}

\definecolor{IEEEblue}{RGB}{0 98 155}
\definecolor{IEEEorange}{RGB}{225 163 0}

\begin{tikzpicture}[spy using outlines={circle, magnification=3.25, size=1.0cm, connect spies}]

\begin{axis}[%
name=boundary,  
width=0.951\figurewidth,
height=\figureheight,
at={(0\figurewidth,0\figurewidth)},
scale only axis,
tick align=inside,
axis line style = thick,	    
xmin=1,
xmax=49.9,
xlabel={Message passing iteration $p$},
ymode = log, 
ylabel style = {yshift=-0.3mm},
ymin=0.3,
ymax=55,
ylabel={$\lightspeed| \toffsetHat{j} - \toffset{j} |$},
ytick=      {0.3, 1, 3,    10,  30,   55},
yticklabels={0.3, 1, 3,   10,   30,   },
grid=none,          
line cap = round,
line join = round,
xmajorgrids,
ymajorgrids,
zmajorgrids,
major grid style={thin,loosely dotted}, 
legend style={legend cell align=left, align=left, draw=white!5!black,fill=white, fill opacity=0.90,font=\scriptsize} 
]

\addplot [color=AP2color, line width=\trajectoryLW, line join=round, line cap=round, 
]
  table[row sep=crcr]{%
1	49.9191666445116\\
2	49.6700424493493\\
3	48.7006122933484\\
4	46.1633250933663\\
5	41.8108657151299\\
6	36.0846017416521\\
7	29.4229673744636\\
8	22.6229907838766\\
9	16.4964581131713\\
10	11.3637841383112\\
11	7.5515008274665\\
12	4.7519737676196\\
13	2.94484835146814\\
14	1.89799689446891\\
15	1.31105807440761\\
16	0.990118661724861\\
17	0.79389193333646\\
18	0.68160183905543\\
19	0.631351876630776\\
20	0.596093453104875\\
21	0.565118981923905\\
22	0.558332479206736\\
23	0.513004701711463\\
24	0.492010466177464\\
25	0.473383274214911\\
26	0.456468303544877\\
27	0.441545873264278\\
28	0.426837318702701\\
29	0.413078364882162\\
30	0.401458707989108\\
31	0.388621205348254\\
32	0.37804959471147\\
33	0.367723509094201\\
34	0.357880133626259\\
35	0.348881315282093\\
36	0.339398426574359\\
37	0.331854030386838\\
38	0.32369109030127\\
39	0.316083578210865\\
40	0.309164574244769\\
41	0.302277338850567\\
42	0.29607437948904\\
43	0.289307394512412\\
44	0.283875745065094\\
45	0.277758526200092\\
46	0.272888861667241\\
47	0.267542378882176\\
48	0.262412959759305\\
49	0.257706427188901\\
50	0.252740999554446\\
51	0.248016828994144\\
};

\addplot [color=AP3color, line width=\trajectoryLW, line join=round, line cap=round, 
]
  table[row sep=crcr]{%
1	49.9472219011334\\
2	49.92274464335\\
3	49.8698879732892\\
4	49.729115140166\\
5	49.3176989468038\\
6	48.2586554864085\\
7	46.3682861639038\\
8	43.3240279628232\\
9	38.94101157909\\
10	33.3039213680998\\
11	26.9321329504808\\
12	20.5335181900364\\
13	14.4648340972912\\
14	9.66922169274923\\
15	6.07316690959728\\
16	3.58025672451424\\
17	2.06590673471324\\
18	1.43422025574049\\
19	1.24876335637944\\
20	1.15827162328814\\
21	1.07882990110185\\
22	0.997973672600176\\
23	0.980298024255189\\
24	0.888441786116442\\
25	0.840581537906841\\
26	0.803308414185566\\
27	0.77011489489047\\
28	0.741400727576149\\
29	0.715285807318537\\
30	0.691474423730499\\
31	0.669164310337863\\
32	0.647245587299609\\
33	0.62778330980252\\
34	0.609997557816544\\
35	0.59349427206592\\
36	0.577230230763982\\
37	0.561825283590336\\
38	0.547486920058805\\
39	0.533509326927044\\
40	0.51926116007444\\
41	0.506400848703966\\
42	0.494455144936307\\
43	0.482541581065345\\
44	0.471560699079202\\
45	0.460693392848918\\
46	0.450239144590342\\
47	0.440607801870245\\
48	0.431828854467431\\
49	0.422771217657368\\
50	0.414221638551352\\
51	0.406457481986782\\
};
\end{axis}

\node[draw,fill=white,inner sep=1pt,below=-0.0em,xshift=-0.0mm,line cap = round, line join = round, line width = 0.5pt]
at (boundary.north) {\footnotesize \color{black}\,c) Clock time offset (\SI{}{\centi\metre})\,};

\end{tikzpicture}%
\vspace{-3mm}
\caption{%
RMSEs 
versus message passing iterations $p$.}
\label{fig:RMSE-vs-iterations}
\end{figure}
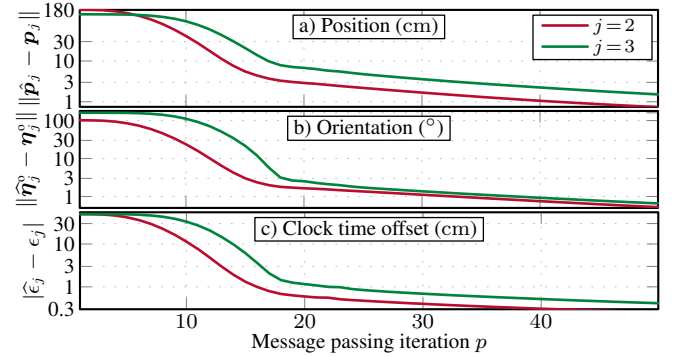
\tikzexternalenable



\section{Conclusions}\label{sec:results}
Our implementation of particle-based loopy \gls{bp} 
infers the positions, orientations, and clock time offsets of multiple agents based on a single bidirectional snapshot of noisy channel observations between each aperture pair.
One promising future research direction is the online phase estimation of distributed arrays, which---together with spatial parameter estimation---enables \acrlong{cjt}. 
This emerging paradigm in distributed \acrshort{mimo} promises to deliver high-rate communications, precise localization, and efficient \acrlong{wpt}.

\appendix

\section{Efficient Implementation}\label{app:implementation}
\noindent The \gls{ml} estimator for the noise variance $\noiseVariance{j}{j'}$ in~\eqref{eq:noiseVarianceHat} is implemented efficiently in a multipath channel (cf.\,\cite[App.\,C]{Deutschmann25OJSP}) and even more efficiently in a \acrlong{los} channel using
\ifthenelse{\equal{\shortPaperVersion}{true}}%
{%
\begin{align}
    \noiseVarianceHat{j}{j'}
    &= \frac{1}{\Nchannel}  ~ \tr\Big( \big(\eye{\Nchannel}\!-\!\nicefrac{\dictEntry{j}{j'}\dictEntry{j}{j'}^\herm}{\Nchannel}\big) 
    \Rhate{j}{j'}\Big) \nonumber\\
    &= \frac{1}{\Nchannel}  ~ \Big( \tr\big( \observation{j}{j'} \observation{j}{j'}^\herm\big) - \frac{1}{\Nchannel}\tr\big( \dictEntry{j}{j'} \dictEntry{j}{j'}^\herm \observation{j}{j'} \observation{j}{j'}^\herm \big)\Big) \nonumber\\
    &= \frac{1}{\Nchannel}  ~ \Big(  \observation{j}{j'}^\herm \observation{j}{j'} - \frac{1}{\Nchannel} \tr\big( \observation{j}{j'}^\herm\dictEntry{j}{j'} ~ \dictEntry{j}{j'}^\herm \observation{j}{j'}  \big)\Big) \nonumber \\
    &= \frac{1}{\Nchannel}  ~   \left\| \observation{j}{j'}\right\|^2 - 
    \underbrace{\frac{1}{\Nchannel^2} \left|  \dictEntry{j}{j'}^\herm \observation{j}{j'}  \right|^2}_{\triangleq |\amplitudeHat{j}{j'}|^2} \, .
    \label{eq:noiseVarianceHatEfficient}
\end{align}
}{%
\begin{align}
    \noiseVarianceHat{j}{j'}
    &= \frac{1}{\Nchannel}  ~ \tr\Big( \big(\eye{\Nchannel}\!-\!\nicefrac{\dictEntry{j}{j'}\dictEntry{j}{j'}^\herm}{\Nchannel}\big) 
    \Rhate{j}{j'}\Big) \nonumber\\
    &= \frac{1}{\Nchannel}  ~ \Big( \tr\big( \observation{j}{j'} \observation{j}{j'}^\herm\big) - \frac{1}{\Nchannel}\tr\big( \dictEntry{j}{j'} \dictEntry{j}{j'}^\herm \observation{j}{j'} \observation{j}{j'}^\herm \big)\Big) \nonumber\\
    &= \frac{1}{\Nchannel}  ~ \Big(  \observation{j}{j'}^\herm \observation{j}{j'} - \frac{1}{\Nchannel} \tr\big( \observation{j}{j'}^\herm\dictEntry{j}{j'} ~ \dictEntry{j}{j'}^\herm \observation{j}{j'}  \big)\Big) \nonumber \\
    &= \frac{1}{\Nchannel}  ~   \left\| \observation{j}{j'}\right\|^2 - 
    \underbrace{\frac{1}{\Nchannel^2} \left|  \dictEntry{j}{j'}^\herm \observation{j}{j'}  \right|^2}_{\triangleq |\amplitudeHat{j}{j'}|^2} \, .
    \label{eq:noiseVarianceHatEfficient}
\end{align}
\vfill\pagebreak
}

\clearpage


\renewcommand{\baselinestretch}{0.01}\small\normalsize 
\makeatletter
\patchcmd{\thebibliography}
  {\settowidth}
  {\setlength{\itemsep}{0.4em}\setlength{\parsep}{0pt}\settowidth}
  {}{}
\makeatother
\bibliographystyle{IEEEbib}
\balance
\bibliography{IEEEabrv,bibliography,bibliographyCISA,refs}

\end{document}